\documentclass[12pt]{article}
\usepackage{jheppub}
\usepackage{mathtools,amssymb,amsthm}
\usepackage[utf8]{inputenc}
\usepackage{enumitem}
\usepackage{xcolor}
\usepackage{amsmath}
\usepackage{caption}
\usepackage{comment}
\usepackage{amssymb}
\usepackage{tikz}
\usepackage[all]{xy}
\usepackage{graphicx}

\usepackage{babel}

\newcommand{\be}{\begin{equation}}
\newcommand{\ee}{\end{equation}}

\global\long\def\bra{\langle}%

\global\long\def\da{\dagger}%

\global\long\def\ml{\mathcal{L}}%

\global\long\def\ket{\rangle}%

\usepackage[bb=dsserif]{mathalpha}
\usepackage{bm}

\def \sec{\begin{section}}
\def \esec{\end{section}}

\def \la {\lambda}

\def \Oc {\mathcal{O}}

\def \Fc {\mathcal{F}}

\def \Lc {\mathcal{L}}

\def \pr {\partial}
\def \ra {\rightarrow}

\def \ee {\mathbb{E}}

\def \beq { \begin{equation}}
\def \eeq {\end{equation}}

\DeclareMathOperator*{\Tr}{Tr}

\def \l {\left}
\def \r {\right}

\def \bra {\langle}
\def \ket {\rangle}

\newcommand \bin[2] {
\begin{pmatrix}
#1 \\
#2
\end{pmatrix}}

\begin{document}
\title{Revisiting Brownian SYK and its Possible Relations to de Sitter}

\author[a]{Alexey Milekhin,}

\affiliation[a]{Institute for Quantum Information and Matter, California Institute of Technology, Pasadena, CA 91125, USA}
\emailAdd{milekhin@caltech.edu}

\author[b]{Jiuci Xu}

\affiliation[b]{Department of Physics, University of California, Santa Barbara, CA 93106, USA}
\emailAdd{Jiuci\_Xu@ucsb.edu}

\abstract{We revisit Brownian Sachdev--Ye--Kitaev model and argue that  it has emergent energy conservation overlooked in the literature before. We solve this model in the double-scaled regime and demonstrate hyperfast scrambling,  exponential decay of correlation functions, bounded spectrum and unexpected factorization of higher-point functions. We comment on how these results are related to de Sitter holography.
}

\maketitle
\section{Introduction}
In the past 25 years we learned a lot about anti-de Sitter (AdS) space holography. However, we live in an expanding Universe and it would be very interesting to formulate the holographic correspondence there. As a first step, one can start from the de Sitter (dS) space. Unlike AdS, dS does not have a natural boundary, so it is not clear where to put the holographic screen. In one of the approaches \cite{Strominger:2001pn,Strominger:2001gp,Anninos:2011ui,Harlow:2011ke,Maldacena:2019cbz}, the dual conformal field theory (CFT) lives at the future/past infinities. However, in this case the dynamical aspects are obscure. Another natural candidate is the cosmological horizon, or to be precise, the associated stretched horizon \cite{Banks:2001px,Banks:2003ta,Susskind:2022bia,Susskind:2022dfz,Susskind:2023hnj}. 
This choice is motivated by the fact that it is the surface of maximal area, so it should have enough degrees of freedom to describe the bulk physics inside the static patch. 
In this approach, however, the gravity remains dynamical on the holographic screen, so it is not clear how to define diff-invariant observables. In lower dimensions the gravity is rigid, so this problem is less severe. A related obstacle is the absence of natural time in dS. Recently this problem was addressed \cite{Chandrasekaran:2022cip} by adding an observer worldline to dS and gravitationally dressing all the observables to it.

Ignoring this issue, it is possible to formulate a number of natural properties, independent of the number of dimensions, which the system living on the stretched horizon must satisfy \cite{Susskind:2021esx,Susskind:2022bia,Susskind:2022dfz,Susskind:2023hnj,Rahman:2022jsf}:
\begin{itemize}
\item The density matrix is maximally mixed \cite{Bousso:2000md,Bousso:2002fq,Chandrasekaran:2022cip}. More precisely, the state of empty dS static patch has the maximal entropy.
\item Despite that \footnote{Of course, having exponentially decaying correlation functions is not unusual. What is unusual is that it is supposed to happen at infinite temperature. For example, in any $1+1$ CFT it is not possible, as the two-point function is fixed by the conformal symmetry to be $\sim (\sinh(\pi(t-x)/\beta))^{2 \Delta}$, so it decay instantaneously at $\beta=0$.} correlation functions exponentially decay at late times \cite{Lin:2022nss,Rahman:2022jsf}
\item Higher-point functions of light operators approximately factorize, assuming that the bulk
theory is weakly coupled
\item The static patch geometry has a time-like Killing vector, so the dual system has a conserved Hamiltonian
\item The energy spectrum is bounded: black holes in dS have a maximal mass.
\item "Hyperfast scrambiling": scrambling time is short, of order 1 in the units of dS radius. 
\end{itemize}

This so-called "hyperfast" scrambling should be contrasted with fast scrambling in usual holographic CFTs, where it happens at times $\sim \log(1/G_N)$. However, one objection to hyperfast scrambling in dS, is that two-point function does not completely decay before $\log(1/G_N)$, so scrambling in the sense of delocalization of information cannot happen prior to that time. We will comment on this more in Section \ref{sec:ds} and in the Conclusion, but we delegate the detailed discussion to a separate paper \cite{AJ}. 

The conjecture of \cite{Susskind:2021esx,Susskind:2022bia,Susskind:2022dfz,Susskind:2023hnj} is that the above properties hold for the so-called double-scaled Sachdev--Ye--Kitaev (DSSYK) model \cite{SachdevYe,kitaevfirsttalk,ms,Polchinski:2016xgd}. 
This conjecture remains to be checked, because DSSYK is complicated, especially at late times and for light operators. \textit{The purpose of this paper is to study the Brownian version of this model and argue that all of the aforementioned properties hold there.}

In the past decade it was realized that black holes are chaotic and 
because of that a lot of their properties are universal. It would be extremely interesting to understand
what is the analogue of this statement for dS.
Brownian DSSYK is too simple to describe all physics of dS, for example correlators only match at late times.
But despite that, we take that all other matching suggest that dS
and Brownian DSSYK maybe, in some sense, in the same "universality class" at late times.

The Hamiltonian of (non-Brownian) SYK model reads as 
\beq
H_{\rm DSSYK} = \sum_{i_1 \dots i_p=1}^N J_{i_1 \dots i_p} \psi_{i_1} \dots \psi_{i_p},
\eeq
with
\beq
\bra (J_{i_1 \dots i_p} )^2 \ket = \frac{N J^2}{2 p^2} \bin{N}{p}^{-1}
\eeq
Operators $\psi_i, \ i=1,\dots,N$ are the standard Majorana fermion operators:
\beq
\{\psi_i, \psi_j\} = 2 \delta_{ij}.
\eeq
The double-scaling regime \cite{Cotler:2016fpe,Berkooz2018Chord,Berkooz2019Towards} corresponds to $N \ra \infty, p \ra \infty$, but
\beq
\label{eq:lambda}
q=\exp(-\la), \ \la = \frac{2 p^2}{N},
\eeq
remaining fixed. 
The Brownian version has exactly the same Hamiltonian,
\beq
H_{\rm BDSSYK}(t) = \sum_{i_1 \dots i_p} J(t)_{i_1 \dots i_p} \psi_{i_1} \dots \psi_{i_p},
\eeq
but now disorder $J(t)$ is a Brownian variable:
\beq
\bra J_{i_1 \dots i_p}(t_1) J_{j_1 \dots j_p}(t_2)  \ket = \delta(t_1-t_2) \delta_{i_1 j_1} \dots \delta_{i_p j_p} 
\frac{N J^2}{2 p^2} \bin{N}{p}^{-1}.
\eeq
It makes this model much simpler. Because the Hamiltonian is time-dependent, it is difficult
to introduce finite temperature in a sensible way. So all the observables we study in this model will be at infinite temperature (that is, maximally mixed density matrix).

The first question which arises is how can this model have energy conservation? For a fixed realization of $J(t)$ it does not, because there is an explicit external source. We will prove that the energy is  conserved after the $J(t)$ average. Precisely, we
show the following Ward-like identity for arbitrary correlation functions:
\beq
\label{eq:energyC}
\pr_t \bra \dots H_{\rm BDSSYK}(t) \dots \ket = 0.
\eeq
This has nothing to do with the double-scaling and this relation holds for arbitrary $N$ and $p$. This property was overlooked in the previous studies of this model.

In our dictionary we identify the disorder variance $J^2$ with the radius of $2+1$ dimensional dS
\beq
\boxed{J^2  = \frac{1}{R_{dS}}},
\eeq
but more importantly, $\lambda$ in DSSYK (eq. (\ref{eq:lambda})) with Newton constant $G_N$ in the bulk:
\beq
\boxed{\lambda = \frac{8 G_N}{R_{dS}}  }.
\eeq
We also show that in Brownian DSSYK, particles have
maximal energy of
\beq
E_{\rm max} = \frac{1}{8G_N},
\eeq
which is a property of $2+1$ dimensional dS.

We also compute higher-point correlation functions, both time-ordered (TOC) and out-of-time 
ordered (OTOC).
For finite $\lambda$, each term in the Hamiltonian mixes $\sim \sqrt{N}$ fermions, so we observe
``hyperfast'' scrambling as expected: the OTOC decays to zero at times of order $1/J^2$. 
However, in this regime
we naively do not expect the factorization of TOC. Interestingly, we do find
that TOC factorises for light operators. This suggests that we can give a bulk interpretation
for these correlation functions.

We also observe the following phenomena in the 4-point function. Suppose we have two types of particles, $V$ and $W$. We found that the decay of $\bra WW \ket$ two-point function slows down in the background of $V$ particle, as can be measured by the TOC $\bra VWWV \ket$. One can think about it as a consequence of infinite temperature, as at infinite temperature the decay rate should be the fastest.
We discuss this phenomena in a separate publication \cite{AJ}. Another non-trivial phenomena is the approximate factorization of higher-point correlation functions: naively at finite $\lambda$ the mean-field analysis is not valid, so one does not expect the usual large $N$ factorization. Surprisingly, we do find that the correlation functions of light operators approximately factorize up to corrections which go as $\lambda$ times mass squared. This is one of the reasons why we associate $\lambda$ with $G_N$.

The rest of the paper is organized as follows.
In Sections \ref{sec:toc} and \ref{sec:otoc} we compute time-ordered and out-of-time ordered 
correlation function. We demonstrate approximate factorization and observe emergent energy conservation.
In Section \ref{sec:ec} we explain the origin of this energy conservation, which holds
even for finite $p,N$. This Section can be read separately. Section \ref{sec:hs} is dedicated
to a Hilbert space interpretation of Brownian chords. Section \ref{sec:ds} reviews our results
in the light of dS physics. We explain similarities as well as differences.

\textbf{Note added:} when this paper was at the finial stages of preparation, ref. \cite{Narovlansky:2023lfz} appeared which also studies the relations between DSSYK and dS.
There are several similarities and differences in our approaches and results. Ref. \cite{Narovlansky:2023lfz}
studies non-Brownian model and relates its correlators to correlators in dS inserted at the podes. The time in DSSYK is identified with the proper time in dS. In this paper we put the correlators on the stretched horizon and identify the time in Brownian DSSYK with the time in the static patch. Interestingly, both papers identify $\lambda=2p^2/N$ with the Newton's constant $G_N$. Because of that, we have an overlap in explaining the dS entropy $\sim 1/G_N$ as accessible entropy in DSSYK, rather than the full entropy (Section \ref{sec:ds}). 
Also another very recent paper \cite{Stanford:2023npy} studied double-scaled Brownian SYK from the scramblon perspective.

\section{Time-ordered correlation functions}
\label{sec:toc}
In this paper we will use the chord technique which greatly simplifies in the Brownian setup. Let us briefly review non-Brownian case. The basic idea is to note that at large $N$ and $p$, composite fermionic operators satisfy a simple commutation relations. 
Let us write down the Hamiltonian as
\beq
H = \sum_\alpha J_{\alpha} \Psi_\alpha, \ \Psi_\alpha = \psi_{\alpha_1} \dots \psi_{\alpha_p}.
\eeq
The matter operators we will be interested in, have a similar form:
\beq
V_\Delta = \sum_{M} C_M \Psi_M, 
\eeq
with $C_M$ being random Gaussian and the size $|M|$ of the multi-index $M$ is large, such that $\Delta=|M|/p$ is kept fixed.
Correlation function of the form
\beq
\Tr H^2 V_\Delta H^2 V_\Delta
\eeq
are easy to average over $J,C$ by doing Wick contractions. One of the terms will have the form
\beq
\label{eq:example}
\sum_{\alpha, \beta, M} \Tr \Psi_\alpha \Psi_\beta \Psi_M \Psi_\alpha \Psi_\beta \Psi_M.
\eeq
Representing $\Tr$ as a circle, the above expression can be drawn as
\begin{center}
\includegraphics{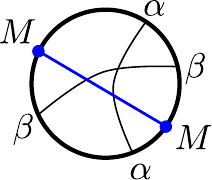}
\end{center}
Each Hamiltonian chord comes with a factor of $J^2/\lambda$, this is just normalization.
The main non-trivial fact is that in the double-scaling limit $\Psi_\alpha$ obey \cite{Berkooz2018Chord,Berkooz2019Towards}
\beq
\Psi_\alpha \Psi_\beta \approx \Psi_\beta \Psi_\alpha \exp( -2|\alpha| |\beta|/N)
\eeq
for generic $\alpha,\beta$. Hence, commuting them past each other in (\ref{eq:example}) yields $q_V^2 q$ with
\beq
q_V \equiv q^{|V|/p} \equiv q^\Delta.
\eeq
The variance of $J$ and $C$ takes care of the combinatoric coefficients, such that one only needs to sum over all possible configurations of chords with the corresponding $q$ weights.

For non-Brownian SYK with time-independent disorder, the chords can be highly non-local in time, resulting in complicated configurations.
In the Brownian case there are only a few chord configurations.
Let us start from 2-point function:
\beq
\bra V_\Delta(t) V_\Delta(0) \ket = \Tr \l(  e^{i \int H dt} V_\Delta e^{-i \int H dt} V_\Delta\r).
\eeq
We can expand each exponent as 
\beq
\sum_k i^k \int_{t_1 < \dots < t_k } dt_1 \dots dt_k \  H(t_1) \dots H(t_k)
\eeq
Since we have a natural forward evolution $e^{-i \int H dt}$ and backward evolution $e^{+i \int H dt}$ it would be convenient to represent the trace as an elongated circle. 
\textit{Now, the key feature of the Brownian case that we only have Wick contractions between $H$ at the same time.}
\begin{figure}
\centering
\includegraphics[scale=0.7]{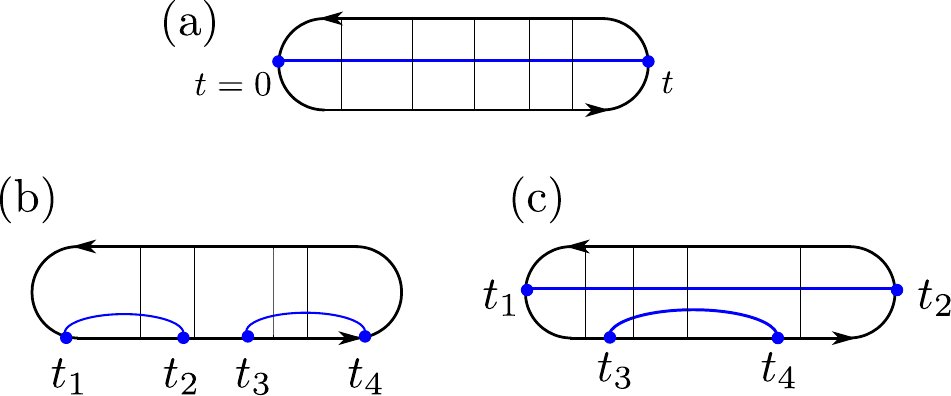}
\caption{(a) Computation of the two-point function using Schwinger--Keldysh contour. Blue chords are matter chords and the black chords represent Hamiltonians.
The same contour can be used to compute time-ordered four-point function. (b) The configuration $\bra V(t_1) V(t_2) W(t_3) W(t_4) \ket$. (c) The configuration $\bra V(t_1) W(t_3) W(t_4) V(t_2) \ket$. }
\label{fig:2pt}
\end{figure}
This way we get two simple types of chords.
First of all, chords
which join forward evolution with forward and backward with backward. They are inserted essentially in the same point, so they never intersect
with anything. But they are important for normalizing the answer: without extra insertions the forward evolution must cancel with the backward evolution. These contact terms produce
\beq
e^{-t J^2/\lambda},
\eeq 
where $t$ is the total length of the contour. 
And then there are the chords which join forward and backward. They are vertical because they must connect points of the same physical time - 
Figure \ref{fig:2pt} (a).
But time-ordered correlators do not intersect. So the answer is 
very simple:
\beq
e^{-t J^2/\lambda }\sum_{k=0}^{+\infty} \frac{J^{2k}}{k!} t^k q_V^k \lambda^{-k} = 
\exp\l( -\frac{J^2}{\lambda}(1-q_V)t \r).
\label{eq:2pt}
\eeq
It exponentially decays at long times.
For light operators $\Delta \la \ra 0$, we get $e^{-\Delta J^2 t}$. We see that we get a finite correlation length at infinite temperature. Also $\lambda$ and "energy" $\Delta$ come together. In Section \ref{sec:ds} we will discuss how it relates to dS.

We can compute higher point time-ordered correlation function as well. For the four-point function in Figure \ref{fig:2pt} (b) the answer completely factorizes:
\beq
\label{eq:4pt_comp}
\bra V(t_1) V(t_2) W(t_3) W(t_4) \ket
= \bra V(t_1) V(t_2) \ket
\bra W(t_3) W(t_4) \ket + \Oc(1/p, 1/N).
\eeq
Such complete factorization might be surprising because we do not expect it in gravity. Perhaps the corresponding connected contribution in gravity is exponentially suppressed at late times, this is why it is not captured in Brownian DSSYK.

For a more complicated kinematics
in Figure \ref{fig:2pt} (c) the answer looks more interesting:
\begin{align}
e^{-(t_2-t_1)/\la} \sum_{k=0}^{+\infty} \frac{1}{k!} J^{2k} \lambda^{-k} (q_V(t_3-t_1)+q_V(t_2-t_4)+q_V q_W(t_4-t_3))^k = 
\nonumber \\
=\exp \l( \frac{J^2}{\la} \l( -(t_2-t_1)(1-q_V) - q_V(t_4-t_3)(1-q_W) \r)  \r).
\end{align}
Hence, 
\beq
\label{eq:4pt_part}
\bra V(t_1)  W(t_3) W(t_4) V(t_2)\ket
= \bra V(t_1) V(t_2) \ket
\underbrace{\exp \l( -\frac{J^2}{\la} \textcolor{red}{q_V}(t_4-t_3)(1-q_W) \r)}_{\text{almost} \ \bra W(t_3) W(t_4) \ket}  + \Oc(1/p, 1/N).
\eeq
Had it not been for the $q_V$ factor (marked in red), the second multiplier would have been
$\bra W(t_3) W(t_4) \ket$. We see that in the presence of $V$ particle, the decay of $W$ slows down. This makes sense: at infinite temperature things decay the fastest, but the presence of $V$ disrupts that. In \cite{AJ} we will argue that a similar thing happens in dS.

Interestingly, for the light operators $\Delta \lambda \ll 1$ we do get factorization, because $q_V$ becomes of order 1: 
\beq
\label{eq:4pt_exp}
\bra V(t_1)  W(t_3) W(t_4) V(t_2)\ket
= \bra V(t_1) V(t_2) \ket 
\bra W(t_3) W(t_4) \ket \l( 1+\lambda 
\Delta_V \Delta_W J^2 (t_4-t_3)  + \Oc(\lambda^2) \r)
\eeq

In large $N$ theories such factorization happens because answers are dominated by a saddle point, and the connected correlation functions are suppressed by $1/N$. In the double-scaled SYK this is not true, because the relevant parameter which controls the semiclassical expansion is $\lambda=2p^2/N$ and it can be of order 1. So in some sense DSSYK is like finite $N$ SYK \cite{douglas_talk}. We see that in the Brownian case things do factorize, but only for the light operators, $\Delta \lambda \ll 1$. It would be interesting to understand this fact on a more intuitive level. 

Another thing to notice is the peculiar kinematics of the final answer: it only depends on the difference of $t_1-t_2$ and $t_3-t_4$. In the ordinary non-Brownian SYK, a similar occurrence arises as a manifestation of energy conservation. Naively, Brownian SYK has external classical noise, so we cannot expect the energy conservation.
 However, one can insert the Hamiltonian in correlation functions, essentially by drawing an extra chord. This way it becomes apparent that the answer does not depend on the time of the insertion:
 For example:
 \beq
 \label{eq:HVV}
 \bra H(t_1) V(0) V(t) \ket = 0, \ t_1 < 0 < t,
 \eeq
\beq
\label{eq:VHV}
\bra V(0) H(t_1) V(t) \ket = -i\frac{J^2 (1-q_V)}{\lambda} \bra V(0) V(t) \ket, \ 0< t_1< t.
\eeq
Hence \footnote{Naively, the energy seems complex. However, in dS for a free massive field of dimension $\Delta$, the two-point function at late times behaves as $e^{- t \Delta /R_{dS}}$, this is why we associate $E_V$ with energy.} we can assign the following energy to the $V$ particle:
\beq
\label{eq:EV}
E_V = J^2\frac{1-q_V}{\lambda}.
\eeq
For light $\Delta \lambda \ll 1$ particles, $E_V \approx J^2 \Delta$. Notice that the energy is bounded from above even if $\Delta$ is large:
\beq
E_V < \frac{J^2}{\lambda}.
\eeq

In the next Section we will explain that the energy conservation in the form of eq. (\ref{eq:energyC}) is the consequence of the form of the SYK Hamiltonian. In particular, it is true for any $N$ and any $p$.

The answers in this Section in the limit $\lambda \ra 0$ can be matched to the non-double-scaled Brownian SYK by computing the usual ladder diagrams. In fact, in this case only a single rung contributes.  

\section{Out-of-time ordered 4-point function}
\label{sec:otoc}

\begin{figure}
    \centering
    \includegraphics[scale=0.7]{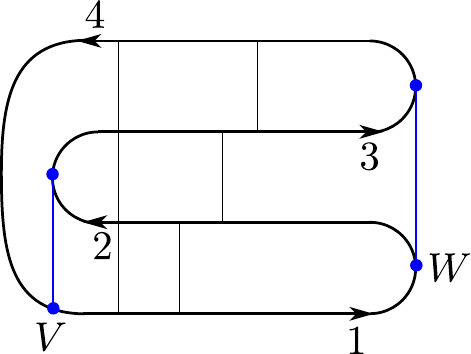}
    \caption{Time-contour which computes the OTOC $\bra V(0) W(t) V(0) W(t) \ket$. Chord $12$ intersects $23$, but $14$ and $34$ do not intersect anything.}
    \label{fig:otoc}
\end{figure}

The computation of the out-of-time ordered correlation function 
\beq
\Fc = \bra V(0) W(t) V(0) W(t) \ket
\eeq
is more complicated, because now spacetime (Hamiltonian) chords can intersect - Figure \ref{fig:otoc}. In total we have 6 possible type of chords, in addition to the matter (blue) ones. 

Combinatorics is much more complicated, but we can easily write down a Schwinger--Dyson equation. 
Lets fix the number of chords to be $n_{12}, n_{34}, n_{13}, n_{14}, n_{23}, n_{24}$ and denote the corresponding contribution by
\beq
\Fc \begin{pmatrix}
n_{12} & n_{34} & n_{13} \\
n_{14} & n_{23} & n_{24}
\end{pmatrix}.
\eeq
This quantity is a sum over all possible chord configurations with the given number of chords.
It is not difficult to see that upon adding an extra chord from the right it obeys the following equation:
\begin{align}
\label{eq:4pt_SD}
\Fc \begin{pmatrix}
n_{12} & n_{34} & n_{13} \\
n_{14} & n_{23} & n_{24}
\end{pmatrix} = 
q_W  \l[ \Fc \begin{pmatrix}
n_{12}-1 & n_{34} & n_{13} \\
n_{14} & n_{23} & n_{24}
\end{pmatrix} + 
\Fc \begin{pmatrix}
n_{12} & n_{34}-1 & n_{13} \\
n_{14} & n_{23} & n_{24}
\end{pmatrix} \r] + \nonumber \\
+ q_V q^{n_{12}+n_{34}+n_{13}+n_{24}} \l[ \Fc \begin{pmatrix}
n_{12} & n_{34} & n_{13} \\
n_{14}-1 & n_{23} & n_{24}
\end{pmatrix} + 
\Fc \begin{pmatrix}
n_{12} & n_{34} & n_{13} \\
n_{14} & n_{23}-1 & n_{24}
\end{pmatrix} \r] - \nonumber \\
-q_V q_W q^{n_{12}+n_{34}+n_{13}+n_{24}-1} \l[
\Fc \begin{pmatrix}
n_{12} & n_{34} & n_{13}-1 \\
n_{14} & n_{23} & n_{24}
\end{pmatrix} +
\Fc \begin{pmatrix}
n_{12} & n_{34} & n_{13} \\
n_{14} & n_{23}-1 & n_{24}-1
\end{pmatrix}
\r].
\end{align}

%%%%%%%%%%%%%%%%%%%% Various limits
It would be convenient to introduce partially resummed $\Fc$:
\beq
\Fc_n = \sum_{n_{12}+n_{13}+\dots n_{34}=n} \Fc 
\begin{pmatrix}
n_{12} & n_{34} & n_{14} \\
n_{13} & n_{24} & n_{23} 
\end{pmatrix}.
\eeq
In this Section we put $J^2=1$.
The actual OTOC is equal to
\beq
\bra VWVW \ket
=e^{-2t /\la}\sum_{n=0}^{+\infty} \Fc_n \frac{t^n}{\lambda^n n!}.
\label{eq:otoc}
\eeq
There are easily soluble limits.
\begin{itemize}
\item $\lambda \ra 0$: conventional large $p$-SYK. Here we expect the usual
fast scrambling instead of hyperfast scrambling. In the limit $\la \ra 0$, with $\Delta$ being a constant number
we can compute first few $\Fc_n$ and notice that
\beq
e^{-2 t /\lambda} \sum_n \Fc_n \frac{t^n}{\lambda^n n!} = 
1 - \Delta^2 \lambda \l(  2 t + \frac{4 t^2}{2} + \frac{8 t^3}{6} + \dots \r) +
\Oc(\lambda^2).
\eeq
Hence the actual OTO 4-point function (\ref{eq:otoc}) behaves as
\beq
1 - \frac{\Delta^2 p^2}{N} (e^{2 t}-1) + \Oc \l( \frac{p^4}{N^2} \r),
\eeq
which coincides with the results of \cite{Stanford:2021bhl}.

\item $\lambda \ra +\infty$: in this limit we do expect hyperfast scrambling because the Hamiltonian
mixes a lot of fermions. In this limit $q \ra 0$, where crossings are suppressed. We can use simple combinatorics to classify all the possible diagrams with the result
\beq
\label{eq:hyperfast}
\Fc = e^{-2t/\lambda} \frac{q_V (1-q_W) e^{2 q_V t/\lambda} - q_W (1-q_V) e^{2 q_W t/\la}}{q_V-q_W}.
\eeq
This is essentially hyperfast scrambling.

\end{itemize}
Interestingly, one can solve the recursion relation (\ref{eq:4pt_SD}). It resembles the recursion relation for $q$-deformed binomial coefficients and by some trial and error one can obtain the following expression:

\begin{equation}\begin{split}
\mathcal{F}\left(\begin{array}{lll}
n_{12} & n_{34} & n_{13} \\
n_{14} & n_{23} & n_{24}
\end{array}\right)  & =(-1)^{n_{13}+n_{24}} q_W^{n_{12}+n_{34}+n_{13}+n_{24}} q_V^{n_{13}+n_{24}+n_{14}+n_{23}} q^{ \frac{\left(n_{13}+n_{24}\right)\left(n_{13}+n_{24}-1\right)}{2}} \\
& \times\left(\begin{array}{c}
n_{12}+n_{34} \\
n_{12}
\end{array}\right)\left(\begin{array}{c}
n_{13}+n_{24} \\
n_{13}
\end{array}\right)\left(\begin{array}{c}
n_{14}+n_{23} \\
n_{14}
\end{array}\right) \frac{\left(q,q\right)_{\sum_{ij}n_{ij}}}{\left(q,q\right)_{n_{12}+n_{34}}\left(q,q\right)_{n_{13}+n_{24}}\left(q,q\right)_{n_{14}+n_{23}}}
\end{split}\end{equation}
where $\sum_{ij} n_{ij} = n_{12}+n_{34}+n_{24}+n_{13}+n_{14}+n_{23}$ and 
$(a,q)_n = \prod_{k=0}^{n-1}(1-a q^k)$.

We can perform a partial resummation over $n_{12}, n_{13}, n_{14}$, by noticing that the answer depends mostly on $n_L \equiv n_{12}+n_{34}, \ n_R=n_{14}+n_{23}, \ n_C = n_{13}+n_{24}$.
This sum will turn the binomial coefficients into $2^{n_C + n_L + n_R}$. Then we can decouple the constraint $n=n_C+n_R+n_L$ by introducing a delta function, $\frac{1}{2\pi }\int_{-\infty}^{+\infty} d\phi e^{i \phi (n-n_L-n_R-n_C)}$ so that we can sum over $n_C,n_R,n_L$ from zero to infinity. 
Then using the identities \footnote{
$$
\sum_{k=0}^{+\infty} \frac{z^k}{(q;q)_k} = \frac{1}{(z;q)_\infty}, \quad 
\sum_{k=0}^{+\infty} \frac{(-1)^k z^k q^{k(k-1)/2}}{(q;q)_k} = (z;q)_\infty
$$
} we get
\beq
\Fc_n = (q;q)_n \frac{1}{2\pi} \int_{-\infty}^{+\infty} d\phi \ e^{i \phi n}
\frac{(2 q_V q_W e^{-i\phi};q)_\infty}{(2q_V e^{-i \phi};q)_\infty
(2q_V e^{-i \phi};q)_\infty}.
\eeq

\section{Intermezzo: energy conservation}
\label{sec:ec}
This Section can be read 
independently from the rest of the paper. The arguments here do not rely on large $N$ or particular space-time dimension.

The goal is to understand when a quantum system with a time-dependent disorder conserves energy.
For each disorder realization the energy obviously is not conserved, but it may become conserved after the disorder averaging.
We will consider the following Hamiltonian:
\beq
\label{eq:genH}
H = H_0 + \sum_\alpha J_\alpha(t) \Psi_\alpha,
\eeq
where $\alpha$, $H_0$ and  $\Psi_\alpha$ are some abstract set indexes and fermionic operators build from $\psi_i$. $J_\alpha(t)$ are classical Gaussian random variables with the
covariance:
\beq
\bra J_\alpha(t_1) J_\beta(t_2) \ket = J^2 \delta(t_1 - t_2) \delta_{\alpha \beta}.
\eeq
The manifestation of energy conservation is more evident in the Hamiltonian picture. In addition, we want a description of the theory where $J(t)$ has been integrated out as implementation of disorder-average.  Such a description is attainable through the Lindbladian formalism. 

\begin{figure}
    
    \begin{minipage}{0.47\textwidth}

    \includegraphics[scale=1.8]{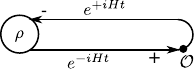}
    
    \end{minipage}
    \begin{minipage}{0.47\textwidth}

    \includegraphics[scale=1.8]{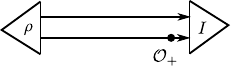}
    
    \end{minipage}
    
    \caption{Left: standard SK contour.
    Right: the same computation but in the doubled Hilbert space formalism.}
    \label{fig:SK}
\end{figure}

Imagine we have a density matrix $\rho$ and we are interested in the expectation value of $\Oc$ after time $t$:
\beq
\label{eq:OcExp}
\bra \Oc(t) \ket = \Tr (\rho e^{+i \int H dt} \Oc e^{-i \int H dt}).
\eeq
Each evolution operator $e^{-i \int H dt}$ can be represented as Feynman path integral leading to the standard Schwinger--Keldysh (SK) contour with $+$ and $-$ sides, Figure \ref{fig:SK} (Left). The time first runs forward on the $+$ contour and then backwards on the $-$ contour.

One can take a different perspective on this picture. Sometimes it is called "third quantization" \footnote{Not be be confused with "third quantization" of baby universes in gravity. } in the condensed matter literature \cite{Prosen_2008,McDonald:2023mbk}. Instead of evolving first forward and then backwards, we can evolve the two sides at the same time. For that we double the Hilbert space and treat the initial density matrix $\rho$ as a state $| \rho \ket$, with $+,-$ labeling the bra and the ket parts. The evolution operator is then called Lindbladian\footnote{Strictly speaking, this is the adjoint of the Lindbladian. The way to to see that, is to recall that Lindbladian has to preserve the trace of the density matrix. The adjoint Lindbladian then has to preserve the maximally mixed state $|I\ket$, which is what we want on the SK contour. } $\Lc$:
\beq
\Lc_{\rm micro} = i H_+ - i H_-.
\eeq
At the "tip" of the SK contour the $+$ has to meet the $-$, so there we insert a maximally mixed state $| I \ket$, between the $+$ and $-$. This way the expectation value (\ref{eq:OcExp}) can be written as
\beq
\bra I | \Oc_+ e^{-\int \Lc_{\rm micro} dt} | \rho \ket,
\eeq
represented by Figure \ref{fig:SK} (Right). 

Let us now discuss the particular case of Hamiltonian (\ref{eq:genH}).
In this language integrating out $J$ is very simple: we have a Gaussian expectation value of $e^{-\int \mathcal{L}_{\rm micro} dt}$ which transforms into $e^{-\mathcal{L} t}$, with

\beq
\Lc = i H_{0,+} - i H_{0,-} + \frac{J^2}{2} \sum_\alpha (\Psi^+_\alpha - \Psi^-_\alpha)^2.
\eeq
This is the effective evolution operator. 
What is the fate of the Hamiltonian operator $H(t)$? 
Without loss of generality, the correlation function of the form
\beq
\Tr \l( \dots H(t) \dots \r)
\eeq
can be embedded into the doubled Hilbert space by putting $H$ on the $+$ side. Then doing Gaussian integral over $J$, the insertion
\beq
H_{0,+} + \sum_\alpha J_\alpha(t) \Psi^+_\alpha,
\eeq
is transformed into
\beq
H_{\rm aver} = H_{+,0} - i J^2 \sum_\alpha \Psi^+_\alpha (\Psi^+_\alpha - \Psi^-_\alpha).
\eeq
The energy will be conserved if $H_{\rm aver}$ commutes with $\mathcal{L}$: $[\Lc, H_{\rm aver}]=0$.
One such example is $H_0=0, \Psi_\alpha^2 = \pm 1$, which is the case of SYK model. In this case $\Lc$ and $H_{\rm aver}$ coincide up to a constant shift.

Despite that the energy is conserved on average, the behavior of other quantities may not follow the general expectations of Hamiltonian systems. For example, in Hamiltonian systems with a finite number of degrees of freedom, correlation functions have Poincare recurrencies. This might not be the case for Lindbladian dynamics, because the evolution operator $e^{-\mathcal{L} t}$  
leads to a monotonic decay. For the case of Brownian SYK it is indeed the case because $\mathcal{L}$ is hermitian.

\section{An algebraic approach towards chords in Brownian DSSYK}

\label{sec:hs}
Similar to \cite{Lin:2022rbf,Lin:2023trc}, in this Section we introduce an auxiliary Hilbert space $\mathcal{H}$ that emerges from the chord rules in Section \ref{sec:ds}. It enables the interpretation of correlation functions, such as \eqref{eq:2pt}, \eqref{eq:4pt_comp}, as transition amplitudes of states that live in $\mathcal{H}$. 

The idea is that we can slice a specific two fold SK contour open at fixed time, with the right half defining a ket and the left half defining a bra.  The correlation function can thus be construed as the inner product of the bra $\bra \text{Left} |$ and the ket $|\text{Right}\ket$. This inner product encompasses a summation over all diagrams featuring open chords entering from the left, connecting to open chords exiting the right. An illustration of this idea is presented below:

\begin{equation}
    \begin{tikzpicture}[baseline={([yshift=-0.1cm]current bounding box.center)},scale=3.5]
        \draw[thick] (0,0) -- (1,0) arc (90: -90: .15) -- (0,-.3) arc (270: 90:.15);
        \draw[thick, blue] plot [smooth] coordinates {(0.2,-0.3) (0.4,-0.2) (0.6,-0.2) (0.8,-0.3)};
        \node at (0.2,-0.3) [circle,fill,blue,inner sep=1.5pt]{};
        \node at (0.8,-.3)  [circle,fill,blue,inner sep=1.5pt]{};
        \draw[thick] (0.1,0) -- (0.1,-0.3);
        \draw[thick] (0.3,0) -- (0.3,-0.3);
        \draw[thick] (0.9,0) -- (0.9,-0.3);
        \draw[thick, dashed, red] (0.6,0.2) -- (0.6,-0.5); 
\end{tikzpicture} \simeq   \Big\bra     \begin{tikzpicture}[baseline={([yshift=-0.1cm]current bounding box.center)},scale=2.5]
    \draw[thick] (0.6,0) -- (0,0) arc (90: 270: .15) -- (0.6,-.3);
    \draw[thick, blue] plot [smooth] coordinates {(0.2,-0.3) (0.25,-0.2) (0.6,-0.15)};
    \draw[thick] (0.1,0) -- (0.1,-0.3);
    \draw[thick] (0.3,0) -- (0.3,-0.3);
    \end{tikzpicture} \Big|  \begin{tikzpicture}[baseline={([yshift=-0.1cm]current bounding box.center)},scale=2.5]
          \draw[thick] (0.6,0) -- (1,0) arc (90: -90: .15) -- (0.6,-.3) ;
       \draw[thick, blue] plot [smooth] coordinates {(0.8,-0.3) (0.75,-0.2) (0.6,-0.15)};
       \draw[thick, black] (0.85, 0) -- (0.85,-0.3);
    \end{tikzpicture} \Big\ket 
\end{equation}
where $\simeq$ means equal up to normalization of the empty diagram, which can be determined through the prescription elaborated in the subsequent discussion. 

We now start with an empty state $|\Omega\ket$ without any chord insertion.  We introduce an operator $H$ that adds a Hamiltonian chord to it. This can be diagrammatically represented as follows:
\begin{equation} \label{eq:def-Omega}
\begin{split}
        |\Omega\ket=\begin{tikzpicture}[baseline={([yshift=-0.1cm]current bounding box.center)},scale=1.5]
    \draw[thick] (0,0) -- (0.2,0) arc (90: -90: .15) -- (0,-.3) ;
    \end{tikzpicture} , \quad
    H|\Omega\ket =\begin{tikzpicture}[baseline={([yshift=-0.1cm]current bounding box.center)},scale=1.5]
    \draw[thick] (0,0) -- (0.2,0) arc (90: -90: .15) -- (0,-.3) ;
    \draw[thick,black] (0.1,0) -- (0.1,-0.3);
    % \node at (0.1,0) [circle,fill,red,inner sep=1pt]{};
    % \node at (0.1,-0.3) [circle,fill,red,inner sep=1pt]{};
    \end{tikzpicture}
\end{split}
\end{equation}
However, we can collapse the Hamiltonian chords in the RHS of \eqref{eq:def-Omega} to a point and ends up with the original state $|\Omega\ket$. That is, $|\Omega\ket$ is invariant with any Hamiltonian insertion $H|\Omega\ket = |\Omega\ket$. Physically this is because the Hamiltonian chords never cross among themselves and one can not distinguish states with different numbers of Hamiltonian chords without matter insertion. They are all equivalent to the state $|\Omega\ket$. Therefore,  the subspace with only Hamiltonian chords contains exactly one state $|\Omega\ket$. \footnote{An alternative way of showing this is to consider the $q\to 0$ limit of the chord algebra $[a,a^\dagger]_q = a a^\dagger - q a^\dagger a =1$, developed in \cite{Berkooz2019Towards} and later associated with a bulk interpretation in \cite{Lin:2022rbf}. The limit eliminates contribution from diagrams with crossings. It has been found in \cite{eremin2008qdeformed} that the limit leads to a completely degenerate spectrum. Instead of describing the algebra as $q\to0$ limit, we intend to view the algebra as emergent from the chord rules associated with the Brownian model in our current work, as we are going to adopt a slicing scheme different from the earlier literature. }

We emphasize that unlike \cite{Lin:2022rbf},  the Hamiltonian chords created above are closed instead of open. The difference origins from the fact that in our setup, we slice the SK contour open at fixed time $t$, and it can never cut any Hamiltonian chords open.  We then associate a state to the slice and construct our auxiliary Hilbert space $\mathcal{H}$ by introducing an inner product that $\bra\Omega|\Omega\ket=1$ and collecting all the matter excitations above it. This can be made precise mathematically through the Gelfand-Naimark-Segal (GNS) construction. In the following, we start to incorporate the matter creators/annihilators into the algebra. 

Unlike Hamiltonian chords, general matter chords intersect with themselves and with the Hamiltonian chord as well. Let's consider matter field of weight $\Delta$. Following the chord rules in Section \ref{sec:toc}, when a matter chord intersects with a Hamiltonian chord, we assign a factor of $q^\Delta$, and when two matter chord intersects each other, we assign a factor of $q^{\Delta^2}$. We then define a matter chord creator $b^\dagger$ as 
\begin{equation}
    b^\dagger |\Omega\ket= |0,\Omega\ket, \quad 
    b^\dagger\begin{tikzpicture}[baseline={([yshift=-0.15cm]current bounding box.center)},scale=1.5]
    \draw[thick] (0,0) -- (0.2,0) arc (90: -90: .15) -- (0,-.3) ;
    \end{tikzpicture}
    =\begin{tikzpicture}[baseline={([yshift=-0.15cm]current bounding box.center)},scale=1.5]
    \draw[thick] (0,0) -- (0.2,0) arc (90: -90: .15) -- (0,-.3) ;
    \draw[thick,blue] plot [smooth] coordinates {(0.12,-0.3) (0.1,-0.2) (0,-0.15)};
    \end{tikzpicture}
\end{equation}
where '0' in $|0,\Omega\ket$ means the number of Hamiltonian chords to the left of the matter chord is 0. The second equation above is a diagrammatic illustration of how $b^\dagger$ acts on $|\Omega\ket$. It's not hard to derive the following commutation relations among $b,b^\dagger$ and $H$:
\begin{subequations}\begin{align}\label{eq:b-commutator}
&[b,b^{\da}]_{r}  =bb^{\da}-rb^{\da}b=1, \quad r=q^{\Delta^2}\\
&H b^{\da} =q^\Delta b^{\da}H, \quad b H=q^\Delta H b  \label{eq:Hb-commutator}
\end{align}\end{subequations}
As an example, the first equation of \eqref{eq:Hb-commutator} can be visualized as:
\begin{equation}
    \begin{tikzpicture}[scale=1.5]
        \draw[thick] (0,0) -- (1,0) arc (90: -90: .15) -- (0,-.3) ;
       \draw[thick, blue] plot [smooth] coordinates {(0.7,-0.3) (0.65,-0.2) (0,-0.15)};
       \draw[thick, black] (0.25, 0) -- (0.25,-0.3);
    \end{tikzpicture} = q^\Delta 
    \begin{tikzpicture}[scale=1.5]
          \draw[thick] (0,0) -- (1,0) arc (90: -90: .15) -- (0,-.3) ;
       \draw[thick, blue] plot [smooth] coordinates {(0.7,-0.3) (0.65,-0.2) (0,-0.15)};
       \draw[thick, black] (0.85, 0) -- (0.85,-0.3);
    \end{tikzpicture}
\end{equation}
A generic state with only one type of matter insertion can be denoted as $|k_1,\cdots, k_m, \Omega\ket$, and we define the action of $b^\dagger$ and $ H$ as
\begin{equation}
    b^\dagger |k_1,\cdots, k_m, \Omega\ket = |0,k_1,\cdots,k_m,\Omega\ket, \quad H|k_1,\cdots ,k_m,\Omega\ket =|k_1+1,\cdots,k_m,\Omega\ket 
\end{equation}
The action of $b$ can then be derived from the commutation relation \eqref{eq:b-commutator} and \eqref{eq:Hb-commutator}.

% }

Now we are ready to evaluate correlation functions with our algebraic formulation. Let's start with the time evolution of $|\Omega\ket$, which we define as  \footnote{$H$ is a two-sided operator, and is related to the Lindbladian generator in section \ref{sec:ec} by $\ml = -H/\lambda $. }
\begin{equation}
    |\Omega(t) \ket =e^{Ht/\lambda} |\Omega\ket = e^{t/\lambda}|\Omega\ket 
\end{equation}
where we have used the fact that $|\Omega\ket$ is invariant under $H$. This can be depicted as the following
\begin{equation}
    |\Omega(t) \ket =
\begin{tikzpicture}[baseline={([yshift=-0.6cm]current bounding box.center)},scale=1.5]
    \draw[thick] (0,0) -- (1,0) arc (90: -90: .15) -- (0,-.3) ;
    \draw[<->] (0,.2) -- node[above=0.05cm] {\textit{t}} (1,.2);
\end{tikzpicture}
\end{equation}
and can be viewed as determining the 'vacuum' correlation function as:
\begin{equation} \label{chord:normalization}
    \bra \Omega | \Omega(T) \ket  =
    \begin{tikzpicture}[scale=1.5]
        \draw[thick] (0,0) -- (1,0) arc (90: -90: .15) -- (0,-.3) arc (270: 90:.15); 
        \draw[<->] (0,.2) -- node[above=0.1] {\textit{T}} (1,.2) ;
    \end{tikzpicture} = e^{\frac{T}{\lambda}}
\end{equation}
Let's move on to the evaluation of two-point function $\bra V(T) V(0) \ket $, this can be evaluated with the following Schwinger-Keldysh contour:
\begin{equation} \label{chord:2pt}
\bra V(T) V(0) \ket =  
    \begin{tikzpicture}[baseline={([yshift=-0.1cm]current bounding box.center)},scale=3.5]
        \draw[thick] (0,0) -- (1,0) arc (90: -90: .15) -- (0,-.3) arc (270: 90:.15);
        \draw[thick, blue] (-0.15,-0.15) -- (1.15,-0.15); 
        \node at (-0.15,-0.15) [circle,fill,blue,inner sep=1.5pt]{};
        \node at (1.15,-0.15) [circle,fill,blue,inner sep=1.5pt]{};
\end{tikzpicture} = \frac{\bra \Omega | b e^{HT/\lambda} b^\dagger | \Omega\ket}{\bra \Omega| e^{HT/\lambda}|\Omega \ket}
\end{equation}
where the second equality above follows from the algebraic prescription. That is, one can slice the contour open and computes it as a transition amplitude for a state with a single open matter chord to itself after evolving for a period of $T$. The evaluation of such an amplitude follows from the commutation relation as:
\begin{equation}
\bra\Omega|be^{HT/\lambda}b^{\da}|\Omega\ket=\sum_{n=0}^{\infty}\frac{T^{n}}{n!\lambda^{n}}\bra\Omega|b H^n b^{\da}|\Omega\ket=e^{Tq^\Delta /\lambda}
\end{equation}
Combined with the normalization \eqref{chord:normalization}, we deduce that
\begin{equation}
\frac{\bra\Omega|be^{HT/\lambda}b^{\da}|\Omega\ket}{\bra\Omega|e^{HT/\lambda}|\Omega\ket}=e^{-\frac{1-q^\Delta}{\lambda}T}
\end{equation}
We can move on to the evaluation of for the time ordered four point function:
\begin{align}
 \bra \textcolor{blue}{V} (T)\textcolor{red}{W}(t_2))\textcolor{red}{W}(t_1) \textcolor{blue}{V}(0)\ket  =  \begin{tikzpicture}[baseline={([yshift=0.1cm]current bounding box.center)},scale=3.5]
        \draw[thick] (0,0) -- (1,0) arc (90: -90: .15) -- (0,-.3) arc (270: 90:.15);
        \draw[thick, blue] (-0.15,-0.15) -- (1.15,-0.15); 
        \node at (-0.15,-0.15) [circle,fill,blue,inner sep=1.5pt]{};
        \node at (1.15,-0.15) [circle,fill,blue,inner sep=1.5pt]{};
        \draw[thick, red] plot [smooth] coordinates {(0.2,-0.3) (0.4,-0.2) (0.6,-0.2) (0.8,-0.3)};
        \node at (0.2,-0.3) [circle,fill,red,inner sep=1.5pt]{};
        \node at (0.8,-.3)  [circle,fill,red,inner sep=1.5pt]{};
        \node at (0.2,-0.4) {$t_2$};
        \node at (0.8,-0.4) {$t_1$};
        \node at (-0.17,-0.38) {$T$};
        \node at (1.17,-0.38) {$0$};
\end{tikzpicture}
\end{align}
where $T>t_2>t_1 > 0$.   In the algebraic language, this corresponds to creating two open matter chords at time $t=0$ and $t=t_1$. The first propagates from $0$ to $T$ and the second propagates from $t_1$ to $t_2$ in the presence of the first. The corresponding amplitude can then be evaluated as
\begin{equation}
\frac{\bra\Omega|b_V e^{H\left(T-t_{2}\right)/\lambda}b_W e^{H\left(t_{2}-t_{1}\right)/\lambda}b_{W}^{\da}e^{Ht_{1}/\lambda}b_{V}^{\da}|\Omega\ket}{\bra \Omega|e^{HT/\lambda}|\Omega\ket}=\exp\left(-\frac{\left(1-q^
{\Delta_V} \right)}{\lambda}T-q^{\Delta_V} \frac{1-q^{\Delta_W}}{\lambda}\left(t_{2}-t_{1}\right)\right)
\end{equation}
This matches the previous result \eqref{eq:4pt_part} with $J$ set to 1.

\section{Comparison with de Sitter}
\label{sec:ds}

\begin{figure}[h!]
    \centering
    \includegraphics{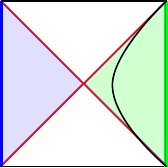}
    \caption{Penrose diagram of global $d$-dimensional dS space.  Two static patches are shown in shaded blue and green. Red is the cosmological horizon. Each point hides an extra $S^{d-2}$. Bold blue and green lines are "antipode" and "pode" where $S^{d-2}$ degenerates to a point. Black line is the stretched horizon.}
    \label{fig:penrose}
\end{figure}

As was argued in \cite{Banks:2001px,Banks:2003ta,Banks:2018ypk,A:2023psv,Susskind:2022bia,Susskind:2022dfz,Susskind:2023hnj,Rahman:2022jsf}, we want to put the holographic screen a few Planck lengths away from the cosmological horizon. That is, on the stretched horizon. There the area, and hence the entropy, is the biggest. 
If we do this in higher dimensions, say in $dS_3$, we can further consider the spherically-symmetric sector to reduce the system to $1+0$ dimensions. 
The metric of empty dS static patch can be written as
\beq
ds^2 = -(1-r^2/R_{dS}^2) dt^2 + \frac{dr^2}{1-r^2/R_{dS}^2} + r^2 d\phi^2.
\eeq
The horizon is at $r=R_{dS}$ and the "pode" is at $r=0$. Time $t$ is the proper time of a static observer at the pode. Penrose diagram is shown in Figure \ref{fig:penrose}.

There are a few very basic expectations, which are all satisfied in the Brownian SYK model:
\begin{itemize}
    \item For the empty static patch the density matrix is maximally mixed.
\end{itemize}
It is indeed the case for the Brownian DSSYK, where the maximally mixed state is very natural to consider.
\begin{itemize}
    \item Despite that, the correlation functions inserted on the stretched horizon, decay exponentially \cite{Rahman:2022jsf}. Namely for operators of dimension $\Delta, \Delta < 1$  and for times $t \gg -R_{dS} \log(G_N/R_{dS})$, we expect
    $\bra V_\Delta(t) V_\Delta(0) \ket \sim e^{-t  \Delta  /R_{dS}}$, where $t$ is the time in the static patch.
    Timescale $\log(1/G_N)$ appears because the operators are inserted at the stretched horizon $r_{sh} - R_{dS} \approx G_N$.
\end{itemize}
This is indeed true in the Brownian DSSYK where $\bra V_\Delta(t) V_\Delta(0) \ket \sim e^{-J^2 \Delta_V t}$. Hence we can identify:
\beq
1/J^2 = R_{dS} ,
\eeq
and more importantly, we identify SYK time with the time $t$ in the static patch, that is, the proper time along the pod. We would like to emphasize that in BDSSYK the exponential answer is exact, whereas in dS it is just a late time approximation. So we cannot talk about a full duality. Perhaps these two models are in the same universality class at late times. 

Naive comparison with free fields propagating in dS suggests that "late" means past $\log(1/G_N)$. This long time scale appears because the field can escape the horizon and propagate through the bulk before falling back in. However, it has also been suggested in the literature \cite{Susskind:2023hnj} that heavy operators like $V_\Delta$ corresponds to the bulk fields confined to the horizon. In such case we do not expect $\log(1/G_N)$ to appear. Instead, the correlator will decay at the timescale of order $R_{dS}$, and we would be able to match it to the Brownian DSSYK at that timescale.
The question about confinement/deconfinement is also related to the scrambling time. It can also clarify whether $V$ is confined to the horizon or not.

\begin{itemize}
    \item In dS the scrambling time is of order $1$.
\end{itemize}
This expectation comes \cite{Susskind:2021esx,Susskind:2022bia,Susskind:2022dfz} from the fact that the dual system already lives on the stretched horizon. In AdS -- black hole spacetime the scrambling time is "long", $\log(1/G_N)$, because it takes this time to fall from the boundary to the stretched horizon. In dS all perturbations are introduced already on the stretched horizon, so naively extra delay timescale is absent. 
In Brownian DSSYK we indeed saw that the scrambling time, as measured by OTOC, is much shorter, of order $R_{dS} \sim 1/J^2$, eq. (\ref{eq:hyperfast}).  

However, we would like to question whether the actual scrambling in dS is hyperfast or not. As discussed above, the two-point function does not decay for a long time, due to the time-dilation at the horizon and the possibility to escape to the bulk. Because of that, the delocalization of information does not happen immediately. \textit{We conjecture that for propagating fields the scrambling is not hyperfast, even if the operators are placed on the stretched horizon}. Instead, it is the usual fast scrambling. We hope to put this conjecture on firmer grounds in a separate paper \cite{AJ}. 
Another way to extract the scrambling time is to compute the switchback time \cite{Susskind:2014jwa} for the holographic complexity. This computation has been done for both empty de Sitter and de Sitter with a black hole \cite{Baiguera:2023tpt,Baiguera:2024xju} and it shows the usual fast scrambling time $\sim \log(1/G_N)$ even for perturbations introduced on the stretched horizon.
We would like to note that if the operators are placed away from the stretched horizon then the scrambling is also fast and such computations have been done in the literature before.
The shockwave computation of OTOC in de Sitter was carried out in \cite{Aalsma:2020aib,Geng:2020kxh} and it was argued that the scrambling is the usual fast scrambling.

\begin{itemize}
    \item The geometry is static (has timelike Killing vector), we must have energy conservation in the dual system. 
\end{itemize}
Surprisingly, this is true for Brownian DSSYK, in the sense of eq. (\ref{eq:energyC}). Although in a very non-trivial way, as explained in the Section \ref{sec:ec}.
\begin{itemize}
    \item For light fields the correlation functions factorize up to $G_N$ corrections.   
\end{itemize}
Again, surprisingly this is true for Brownian DSSYK, where depending on the kinematics, 4-point (or higher) functions either factorize completely (eq. (\ref{eq:4pt_comp})), or approximately (eqns. (\ref{eq:4pt_part}) , (\ref{eq:4pt_exp})).  The rate of non-factorization is governed by $\Delta \lambda$. So we should anticipate that $\lambda \sim G_N$.
\begin{itemize}
    \item A well-known fact that black holes have maximal mass in dS, which goes as 
    $M_{\rm max} \sim \frac{1}{G_N^{d-2}}$ for the case of $d-$dimensional $dS$.
\end{itemize}
By inserting the Hamiltonian operator into the correlation functions (\ref{eq:HVV}), (\ref{eq:VHV}) we argued that an $V$ of dimension $\Delta_V$ produces a particle of energy (eq. (\ref{eq:EV}))
\beq
E_V = J^2 \frac{1-e^{-\lambda \Delta_V}}{\lambda}.
\eeq
As the dimension grows, the energy stays bounded by $J^2/\lambda$. Specifically, for 1+2 dS, $E_{max} = (8 G_N)^{-1}$, hence 
\beq
\lambda =  8 G_N/R_{dS}.
\eeq

\begin{itemize}
    \item dS static patch horizon has entropy $\sim 1/G_N$.
\end{itemize}
At the first sight this is inconsistent with the SYK answer: in SYK the entropy at infinite temperature is $N \log{2}/2$, whereas $1/G_N \sim 1/\lambda \sim N/p^2$.
 In non-Brownian DSSYK case the resolution might come from the fact that there is entropy even at zero temperature \cite{ms},
\beq
S_{T=0} = \frac{N}{2}\log 2 - \frac{\pi^2 N}{4p^2} + \Oc(N/p^3), 
\eeq
so not all of this $N \log(2)/2$ is accessible. The difference in entropies
\beq
\label{eq:deltaS}
\Delta S = S_{T = \infty} - S_{T=0} = \frac{\pi^2 N}{4p^2}
\eeq
scales as $1/\lambda$ as expected.
In the Brownian case it is not clear how to resolve this tension because it is not clear how to go away from the infinite temperature case.

Interestingly, even the above interpretation is missing a numerical factor if we try to compare with entropy of $dS_3$. In that case, the area formula yields
\beq
S_{GH} = \frac{2 \pi R_{dS}}{4 G_N},
\eeq
so the ratio with $\Delta S$ in eq. (\ref{eq:deltaS}) is
\beq
\frac{S_{GH}}{\Delta S} = \frac{8}{\pi}.
\eeq

Notice that in the double-scaling limit, the piece $N \log{2}/2$ is infinite. Perhaps it can be related to observation of \cite{Goheer:2002vf} that dS symmetries are not compatible with finite entropy.

\section{Conclusion}

In this paper we studied Brownian SYK model. Despite its simplicity, it has a lot of unexpected properties. First of all, for any $N$ and $p$ there is an emergent energy conservation, despite the fact that the model has time-dependent disorder. We solved this model in the double-scaling regime and found a number of features which match with dS static patch physics. Namely, exponential decay of correlation function, bounded spectrum, hyperfast scrambling and most surprisingly the approximate factorization of higher-point correlation functions. Also Brownian SYK automatically has infinite temperature. In this matching, we associated SYK time with the dS static patch time (the proper time of a time-like inertial observer sitting at the pod of the dS static patch).
Brownian SYK does not reproduce the full 2-point function of free fields in dS, and it is not clear how to fix this mismatch. What may be true is that Brownian double-scaled SYK captures the \textit{leading} late-time physics of dS static patch. Such interpretation could also explain the complete factorization of the time-ordered four-point function (\ref{eq:4pt_comp}) in a particular kinematics: perhaps the connected piece has an extra exponential suppression in time. It would be interesting to study this directly in the bulk.

Another thing which does not obviously match is the entropy. SYK entropy seems to be parametrically bigger than the area of the cosmological horizon. For non-Brownian SYK we argue that this mismatch can be explained (modulo a numerical factor) by taking a difference between infinite temperature entropy and zero temperature entropy, thus associating dS entropy with the accessible entropy. In the Brownian case it is not clear how to do it because it is not clear how to go away from the maximally mixed state. A more general question is how to define a sensible Hilbert space for the Brownian SYK. 
Naively, we can introduce a fixed density matrix $\rho$. However, it will not affect correlation functions we computed, because none of the Hamiltonian chords will attach to $\rho$. In order for this to happen, $\rho$ has to be correlated with disorder $J(t)$ at later times, which seems unphysical. Despite that, we saw that higher-point correlation functions like $\bra W V V W \ket$ can be given the interpretation of $V$ particle propagating in the background of $W$ particle. Moreover in Section \ref{sec:hs} we discussed the "chord Hilbert space". This makes us hopeful that it is possible to introduce states in a sensible way.

The energy conservation we found might be interesting from the condensed matter perspective. In recent years, random quantum circuits (RQC) \cite{Nahum_2018,von_Keyserlingk_2018,Fisher:2022qey} have provided a rich playground for studying the dynamics of entanglement and measurement-induced phase transitions \cite{Szyniszewski:2019pfo,Li:2018mcv,Li:2019zju,Skinner:2018tjl,Milekhin:2022bzx}. However, RQCs suffer from the lack of energy conservation. Brownian SYK is a unique example where the energy is conserved. It can be thought of as RQC, because at each timestep the evolution operator $e^{-i H \Delta t}$ is not correlated with other ones due to the disorder $J(t)$.
Despite that the energy is conserved on average, some other properties of Hamiltonian systems may not hold. For example, two- and four-point functions decay exactly to zero, with no Poincare recurrencies. 
It would be interesting to investigate whether this "fake" energy conservation impacts other physical properties, such as entanglement and transport. For that one can analyze coupled SYK models or the SYK chain. 
Coupled SYK models exhibit various interesting phenomena, such as spontaneous symmetry breaking \cite{Kim:2019upg,Klebanov:2020kck} and the absence of Schwarzian dominance \cite{Milekhin:2021cou}. However, solving them in the double-scaling regime proved to be very challenging. Perhaps some progress can be obtained in the Brownian case.

Finally, we would like to draw attention to two physical phenomena which goes beyond SYK and which we hope to discuss elsewhere \cite{AJ}.
The first is related to the decay rate of two-point functions in different states. One intuitive statement, which nonetheless is hard to prove is that at infinite temperature correlations decay the fastest. This can be easily seen for large $p$ SYK, $1+1$ CFTs and we also saw this in Brownian SYK, if we interpret the four-point function $\bra V W W V\ket$ as $W$ particle propagating in the background of $V$ particle, eq. (\ref{eq:4pt_part}). By studying the correlation functions of matter fields in dS-black hole geometry one can also see that the two-point function decays the fastest in empty dS. This supports the idea that empty dS has infinite temperature.

The second observation concerns the scrambling time in dS. If the holographic screen is located outside the stretched horizon, then it is natural to expect that the scrambling time is at least $\log(1/G_N)$ because it takes this time to reach the stretched horizon. However, if the holographic screen coincides with the stretched horizon, it is not obvious what happens. Direct bulk computation can be complicated because it involves finding $G_N$ correction to the matter four-point function, which might depend on the precise gravitational dressing of the observables and how one fixes the position of the holographic screen. However, we can try to constrain scrambling by the behavior of two-point function, which is easy to compute. Intuitively, scrambling means delocalization of information, so all two-point functions of local observables should decay to zero before the scrambling can happen. For example, in Brownian SYK at large $\lambda$ we saw that two-point function and four-point OTOC function decay on the same timescale. If we put the holographic screen on the stretched horizon, then the two-point function does not decay for a long time (or order $\log(1/G_N)$) because the excitation can fall into the bulk towards the pode. This suggest that even in this case the scrambling is fast (taking time of order $\log(1/G_N)$) rather than hyperfast (taking time of order $1$). We hope to formalise this argument in a future publication. The absence of hyperfast scrambling will eliminate the need for finite $\lambda$ in SYK. This is a good thing because we associate $\lambda$ with $G_N$.

\section*{Acknowledgement}
We would like to thank Ying~Zhao for the
collaboration at the early stages and numerous illuminating discussions.
We are grateful to Ahmed~Almheiri, Elena~Caceres, Xi~Dong,  Akash~Goel, Alexander~Gorsky, Alexei~Kitaev, Igor~Klebanov, Juan~Maldacena, Donald~Marolf, Emil~Martinec, Mark~Mezei, Vladimir~Narovlansky, John~Preskill, Edgar~Shaghoulian, Thomas~Schuster, Eva~Silverstein for comments and Fedor~Popov for discussions and feedback on the manuscript. 
AM also thanks Cory~King for moral support.

AM acknowledges funding provided by the Simons Foundation, the DOE QuantISED program (DE-SC0018407), and the Air Force Office of Scientific Research (FA9550-19-1-0360). The Institute for Quantum Information and Matter is an NSF Physics Frontiers Center. AM was also supported by the Simons Foundation under grant 376205. J.X. was on the MURI grant and was supported in part
by the U.S. Department of Energy under Grant No. DE-SC0023275. This material is
based upon work supported by the Air Force Office of Scientific Research under award
number FA9550-19-1-0360.

\bibliographystyle{jhep}

\bibliography{refs}

\end{document}